\documentclass[twocolumn,english,aps,prl,groupedaddress,showpacs]{revtex4}
\usepackage[T1]{fontenc}
\usepackage[latin9]{inputenc}
\usepackage{textcomp}
\usepackage{graphicx}
\usepackage{times}

\makeatletter
\@ifundefined{textcolor}{}
{%
 \definecolor{BLACK}{gray}{0}
 \definecolor{WHITE}{gray}{1}
 \definecolor{RED}{rgb}{1,0,0}
 \definecolor{GREEN}{rgb}{0,1,0}
 \definecolor{BLUE}{rgb}{0,0,1}
 \definecolor{CYAN}{cmyk}{1,0,0,0}
 \definecolor{MAGENTA}{cmyk}{0,1,0,0}
 \definecolor{YELLOW}{cmyk}{0,0,1,0}
 }

\makeatother

\usepackage{babel}
\begin{document}

\title{SKATING ON A FILM OF AIR: DROPS IMPACTING ON A SURFACE}

\author{John M. Kolinski$^{1}$, Shmuel M. Rubinstein$^{1}$$^{,}$$^{2}$,
Shreyas Mandre$^{3}$, Michael P. Brenner$^{1}$, David A. Weitz$^{1}$$^{,}$$^{2}$,
and L. Mahadevan$^{1}$$^{,}$$^{2}$}

\affiliation{$^{1}$School of Engineering and Applied Sciences, Harvard University,
Cambridge, Massachusetts 02138, USA\\
 $^{2}$Department of Physics, Harvard University, Cambridge, Massachusetts
02138, USA\\
 $^{3}$Division of Engineering, Brown University, 182 Hope St, Providence,
RI 02912 (USA)\\
}

\date{\today}
\begin{abstract}
Drops impacting on a surface are ubiquitous in our everyday experience.
This impact is understood within a commonly accepted hydrodynamic
picture: it is initiated by a rapid shock and a subsequent ejection
of a sheet leading to beautiful splashing patterns. However, this
picture ignores the essential role of the air that is trapped between
the impacting drop and the surface. Here we describe a new imaging
modality that is sensitive to the behavior right at the surface.
We show that a very thin film of air, only a few tens of nanometers
thick, remains trapped between the falling drop and the surface as
the drop spreads. The thin film of air serves to lubricate the drop
enabling the fluid to skate on the air film laterally outward at surprisingly
high velocities, consistent with theoretical predictions. Eventually
this thin film of air must break down as the fluid wets the surface.
We suggest that this occurs in a spinodal-like fashion, and causes
a very rapid spreading of a wetting front outwards; simultaneously
the wetting fluid spreads inward much more slowly, trapping a bubble
of air within the drop. Our results show that the dynamics of impacting
drops are much more complex than previously thought and exhibit a
rich array of unexpected phenomena that require rethinking classical
paradigms.
\end{abstract}

\maketitle
Raindrops splashing on a car window, inkjets printing on a sheet of
paper and the dripping faucet in the kitchen, are all everyday experiences
which depend on the impact of drops of fluid on a surface. As familiar
as these phenomena are, the impact of a drop of fluid on a surface
is, in fact, quite complex \cite{Mani2010_1,Yarin2006,Courbin2009_1,Schroll2010_1}.
Particularly stunning are the beautiful splashing patterns that often
occur \cite{Worthington1876_1,Josserand2005_1,Bird2008_1}; our understanding
of these is predicated on very rapid impact followed by a shockwave
as the fluid bounces back from the surface \cite{Lesser1981_1,Lesser1983_1}.
However, before contact can occur, the drop must first drain the air
separating it from the surface. Indeed, experimental studies showing
the suppression of splashing at reduced ambient pressure underscore
the importance of the air \cite{Xu2005_1,Driscoll2010_1,Rein2008_1,Mani2010_1,Mandre2009_1}.
Recent theoretical calculations suggest that, even at moderate impact
velocities, the air fails to drain and is instead compressed, deforming
and flattening the bottom of the drop while serving as a thin cushion
of air a few tens of nanometers thick to lubricate the spread of the
drop \cite{Mani2010_1,Mandre2009_1}, and leading to the eventual
formation of a trapped bubble of air within the drop \cite{Thoroddsen2005_1,Thoroddsen2002_1}.
However, the initial stages of impact occur over diminutive length
scales and fleeting time scales, and the very existence of this thin
film of air remains controversial; indeed, this film has never been
directly observed. Moreover, the mechanisms leading the breakup of
this film and the ultimate wetting of the surface have never even
been considered. Testing these ideas requires direct observations
of the impacting interface; however, this demands development of new
experimental methods to attain the requisite spatial and temporal
resolution. 
\begin{figure}
\includegraphics[width=1\linewidth]{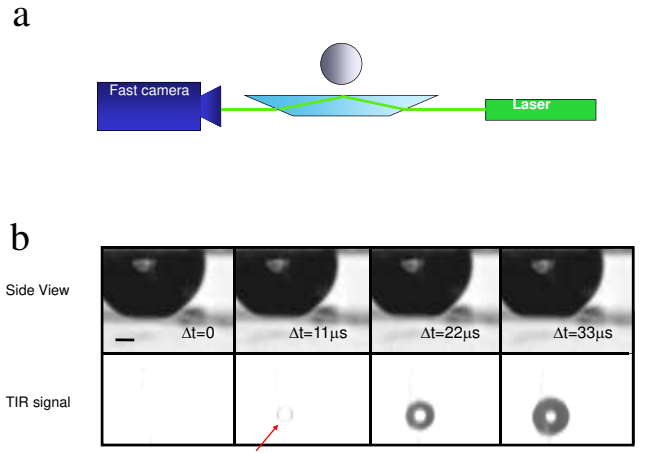} \caption{: Experimental setup. (a) Schematic of TIR microscopy. b. Four typical
images of a 2.6-mm-diameter drop, falling from $H$=21cm\, viewed
simultaneously from the side and with TIR.}
\label{fig1o}
\end{figure}

In this letter, we describe direct measurements of the initial contact
dynamics of a drop impacting a dry glass surface. To visualize the
impact we image from below rather than from the side; to discern the
very thin film we combine total internal reflection (TIR) microscopy\cite{Rubinstein2004_1}
with a novel virtual frame technique (VFT). We directly observe a
thin film of air that initially separates the liquid from the surface
enabling much more rapid lateral spreading of the drop providing striking
confirmation of the theoretical predictions \cite{Mani2010_1}. However,
we also observe a complex sequence of events that leads to the rupture
of the film and ultimate contact of the liquid with the solid surface;
the initially smooth air film breaks-up as discrete holes are formed
and are filled by the liquid. These holes rapidly spread and coalesce
into a ring of wet surface surrounding a trapped bubble of air.

To observe the thin film of air, we illuminate the top surface of an
optically smooth dove prism (BK7 glass) with collimated light incident
from below at an angle greater than the critical angle for total internal
reflection at a glass-air interface but smaller than that at the glass-liquid
interface. The reflected light is imaged with a fast camera, as shown
schematically in fig. 1a. The light reflected from each point of the
interface, $I_r(x,y)$, depends exponentially on the separation between
the impacting fluid and the solid surface, with a characteristic decay
length that depends on the angle of incidence and is of order of 50nm;
as the separation decreases further of the incident light is no longer
fully totally internally reflecting and Ir decreases. This directly
probes the thin film of air. We illustrate this using a 1.3-mm-radius
isopropanol (IPA) drop falling from an initial height $H$=21cm. When
the drop is far from the surface the illuminating beam is totally
internally reflected and nothing is observed as shown in fig.1b;
we define this as $t$=0. However, as the separation between the drop
and the solid surface becomes comparable to decay length of the evanescent
field some of the incident light is no longer totally internally reflected
and $I_r$ decreases; thus, a faint ring is observed as the impact dynamics
begin, at $t$ =11sec. In this case the fluid is not actually wetting
the surface; instead the drop is supported by a thin layer of air.
When wetting finally occurs, there is no longer any totally internally
reflected light and a dark ring is observed, at $t$=22$\mu$sec. As
the drop continues to impinge on the surface the ring of wetting fluid
grows both in the outward and inward directions, as shown for $t$ =33$\mu$sec.
\begin{figure}
\includegraphics[width=0.9\linewidth]{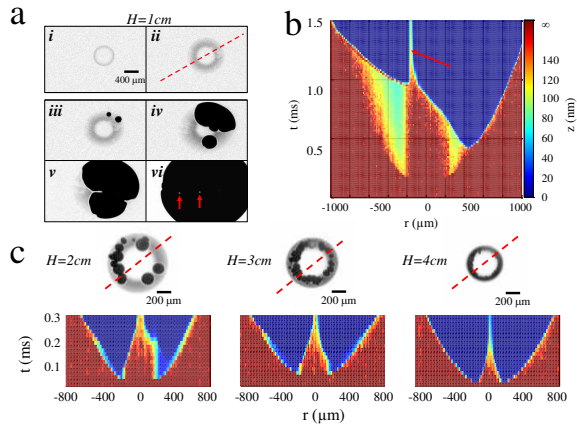} \caption{The behavior of the thin air film separating the impacting drop from
the surface. (a) Six TIR snapshots of a drop, released from $H$=1 cm
illustrating the thin film of air and the impact dynamics. The two
bubbles remaining in the drop are indicated by the arrows in iv. b)
The impact dynamics along the cut shown by the dashed line in (a ii).
The height is indicated by the color. The arrow indicated one of the
bubbles that remains trapped in the liquid c) The TIR images and time
evolution of the air films along the dashed lines for $H$=2, 3 and 4
cm.}

\label{fig2o}
\end{figure}

To elucidate the impact dynamics we explore the behaviour of drops
falling from different initial heights. For $H$=1cm, we first detect
the drop when a thin ring appears with an inner diameter of about
500 $\mu$m, as shown in fig. 2a.i. The outer dimension of the ring
grows rapidly as the drop falls, with an outwards velocity of \textasciitilde{}1
m/s, comparable to the impact velocity of 0.44 m/s, as shown in fig.
2a.ii. However, the fluid does not actually contact the surface; instead,
the fluid spreads on a film of air only \textasciitilde{}100 nm thick.
To visualize the dynamics we take a cut through the image at the location
shown by the dashed red line in fig. 2a.ii, convert the measured intensity
to separation and plot the time evolution, using colour to denote
the height, as shown in the 2D graph in fig. 2b. The first 500 \textmu{}s
clearly show the formation of the layer of air as the drop spreads
before the liquid contacts the surface. The liquid does not spread
inwards, as seen by the boundaries of the thin film, denoted by the
central red region; this reflects the pocket of air which ultimately
becomes a bubble trapped in the drop.

While the layer of air is clearly responsible for decelerating the drop, it cannot retain the separation
of the fluid and surface indefinitely; ultimately, the thin film of
air becomes unstable and contact occurs. Initially, two small dark
spots appear in the film when the liquid fully contacts the surface,
as shown in fig. 2a.iii. These are denoted by the dark blue region
at $t$\textasciitilde{}0.8msec in fig. 2b. As these spots grow, other
spots appear, as the film of air breaks down, as shown in fig. 2a.iv.
These liquid wetting fronts spread rapidly, wetting the surface at
a velocity of 1.5m/s, comparable to that of the liquid spreading on
the thin film of air. Interestingly, there is a thin line of air at
the front of the spreading fluid where the air film becomes thicker
as the air is pushed by the advancing wetting front, as shown by the
white region leading the edge of the black wetting front. Ultimately
two small air bubbles remain, displaced from the center of the drop,
as shown by the arrows in fig. 2a.vi and fig. 2b.

Similar dynamics persist as the initial height of the drop is increased: the drop is
again decelerated by a thin annulus of air with a thicker pocket in
the middle; however, the thickness of the film of air also decreases,
becoming of order 10nm for a drop height of 4cm. As $H$ increases the
initial size of the inner air pocket also decreases; moreover, the
time during which its size remains constant is also reduced. Similarly,
the thin film of air is only clearly observed over a much smaller
region, prior to complete contact. For example, for $H$ = 3cm, the air
film is \textasciitilde{} 20 nm thick and is already only partly observed
at the outer edges of the annulus, as shown by the 2D graph and confirmed
by the snapshot (fig. 2c). As we increase the initial drop height
to 4cm, contact appears to occur around the full ring more rapidly
than our frame rate of 60 kHz; however, even here the initial wetting
is discontinuous, occurring in numerous discrete points as indicated
by the rough texture of the inside of the ring. Thus, the drop is
decelerated by an even thinner film which then breaks up at discrete
locations. As we increase $H$ further, we no longer have sufficient
temporal resolution to observe the initial film of air.

To overcome this inherent limitations imposed by even the highest speed camera,
we introduce a new imaging method, exploiting the fact that the intensity
will change from completely bright to completely dark for a very small
change in the liquid-solid separation. We exploit this nearly binary
contrast by increasing the camera exposure time to integrate over
times longer than the characteristic dynamics. This is illustrated
schematically for a wetting front moving in one dimension in fig.
3a, using a composite image, which reflects the sum of the individual
images at each time. The over-exposed image displays a linear black
to white gradient; this is essentially the sum of a series of individual
virtual frames, which can be recovered by taking consecutive thresholds.
We therefore call this method the virtual frame technique (VFT). The
temporal resolution is determined by the dynamic range of the camera;
thus, using a camera with 14-bit dynamic range, and an exposure time
of 100 $\mu$s the VFT allows us to resolve dynamics as short as 6
ns! This temporal resolution can be further improved by exploiting
the gamma correction, which provides the camera an optional nonlinear
integration time, and is particularly useful for isolating dynamics
of accelerating fronts. Moreover, with VFT, the full spatial resolution
of the camera is preserved. Thus, the VFT provides a combination of
spatial and temporal resolution that is much greater than for any
camera available \cite{Supp}.

We employ the VFT to study the impact dynamics of drops released from initial heights ranging from
1cm to 50cm. For all $H$, the integrated image is disk shaped with a
darker ring where contact first occurs, a bright white spot in the
middle where the air bubble remains, and an evolution from black to
gray to white moving outwards where wetting has not yet occurred,
as shown in fig. 3b. For $H$=2cm there are pronounced features in the
image which are not observed for larger values of $H$, where the images
are more symmetric. These features reflect the non-uniform nature
of the initial wetting, consistent with the images in fig 2c. 

\begin{figure}
\protect\includegraphics[width=0.9\linewidth]{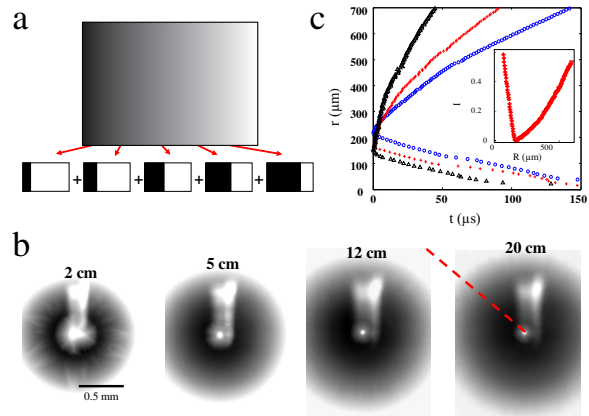}\caption{\label{figure 3:}Virtual Frame Technique (VFT). (a) 1D schematic
demonstrating the concept. Individual binary images, below, are integrated
to yield the total gray-scale image. The gray scale can be interpreted
to yield the time evolution. (b) Four VFT images taken different values
of $H$. Each image exhibits a square-shaped whiter region through the
top center resulting from spurious reflections in the beam path; they
are ignored in our analysis. (c) The intensity is converted to time
and azimuthally averaged around the impact center. The distance of
the wetting fronts from the center are plotted as a function of time
for three typical experiments with $H$=26, 126, 456 mm for blue circles,
red pluses and black triangles respectively. }
\label{fig3o}
\end{figure}

To quantify the VFT data, we measure intensity as a function of radial
distance along the dashed line shown in fig. 3b, and plot the results
in the inset of fig. 3c. The intensity data are converted to time
to obtain the temporal evolution of the front, which is shown for
several values of $H$ in fig. 3c. The lower branch of each curve reflects
the inward-travelling front as the ring closes to entrap the bubble
in the middle of the drop; the upper branch of each curve reflects
the outward-travelling front as the falling drop spreads. The point
where the two meet is the radial distance at which contact first occurs,
$R_0$; this is a decreasing function of initial height, as shown in fig.
4a, and the radial contact disc size exhibits a power-law dependence
on $H$, with an exponent of 1/6, consistent with theoretical predictions\cite{Mani2010_1,Supp},
 as shown in the inset.

To explore the initial dynamics of the wetting associated with the rupture or break down of the air
cushion, we numerically calculate the local instantaneous velocity
and plot its magnitude as a function of radial position, $r$. The inward-moving
velocity is constant, propagating at approximately 1.3 m/s; by contrast,
the outward-moving velocity decreases as $1/r$, and can exhibit remarkably
high values, as large as \textasciitilde{}70 m/s, as shown in fig.
4b. Surprisingly, the velocity of the inward-moving front is independent
of $H$; by contrast the maximum velocity of the outward-moving front
increases strongly with $H$, as shown in fig. 4c. The maximum velocity
is nearly an order of magnitude greater than the capillary velocity
for IPA, $\gamma/\mu$ \ensuremath{\approx} 10 m/s. 

\begin{figure}
\protect\includegraphics[width=0.9\linewidth]{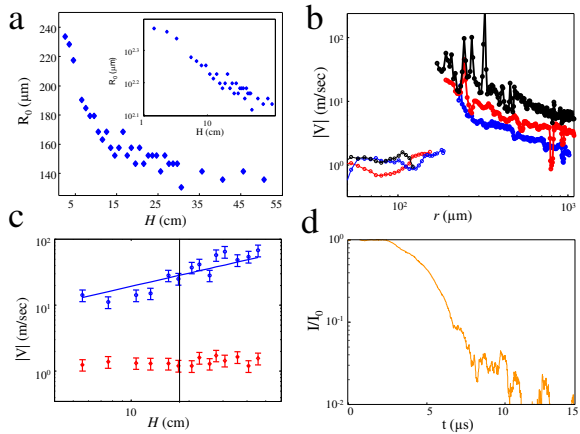}\caption{\label{figure 4:}The initial dynamics of the wetting. (a) $R_0$, as a function of $H$ (inset) same as main figure in log scale. (b) The inwards (solid circles) and outwards (open circles) velocity of the spreading liquid for $H$=26, 126, 456 mm corresponding to blue, red and black respectively. (c) Peak velocities for the outwards (blue circles) and inwards (red circles) fronts. Blue curve is the theoretically predicted initial outwards spreading velocity \cite{Supp}. The dashed line indicates the threshold height above which splashing is observed. (d) A photo diode trace acquired at 100MHz measu ring the intensity of the reflected light directly underneath the thin air film. }
\label{fig4o}
\end{figure}

Since such large velocities are unlikely for a fluid spreading directly
in contact with a surface, these observations suggest that the fluid
is still spreading on a thin film of air. Indeed, such high velocities
are predicted theoretically\cite{Supp}, but only with the explicit
assumption that the spreading occurs over a film of air, as indicated
by the excellent agreement between the calculated behaviour, shown
by the solid line, and the data in figure 4c. Although the VFT assumes
nearly binary data, the resulting virtual frames will be practically
indistinguishable for a simple dry-wet transition and an extremely
short lasting air film which is followed immediately by a wetting
front. To determine whether there is indeed a thin film of air, we
measure the intensity at a point using a photodiode operating at 100
MHz. The intensity initially drops rapidly, but then remains nearly
constant for a period of \textasciitilde{}2.5s, whereupon it again
rapidly drops, as shown in fig. 4d. This observation is consistent
with a picture in which the fluid spreads on a thin film of air of
order 10nm thick; this is trailed closely by a wetting front that
rapidly expands due to the breakdown of the air film.

Our results directly demonstrate the existence of a thin film of air over which the liquid spreads; this provides striking confirmation of the theoretical prediction\cite{Mani2010_1,Mandre2009_1}.
In addition, our results reveal that qualitatively new phenomena occur as the thin film of air becomes unstable; simultaneously breaking
down at many discrete locations, leading to wetting patches that grow
and coalesce to fully wet the surface. For a perfectly wetting fluid
such as IPA on glass, a thin film of air behaves as does a poor solvent;
it cannot remain stable and van der Waals forces will cause it to
de-wet the surface through a nucleation or spinodal-like process \cite{Reiter1992_1,DeGennes2004_1};
indeed fig. 2a.ii is reminiscent of the patterns observed in such
processes. For spinodal de-wetting the rate of film breakup depends
strongly on its thickness \cite{DeGennes2004_1}; for example, a 10nm
thick air film will remain stable for no longer than one microsecond. Thus, rupturing occurs simultaneously at many discrete locations; this leads to small wetting patches that grow and coalesce to fully cover the surface, thereby very rapidly following the advancing fluid front. This gives the appearance of a single contact line moving at the same velocity as the fluid, much faster than the calculated capillary velocity.

Using a novel experimental modality that visualizes the falling drop
from below rather than from the side, we identify a thin film of air
that initially separates the liquid from the surface. Eventually,
however, spinodal-like dewetting of the air film always leads to its
breakup and complete contact of the surface by the fluid. The rate
at which contact occurs depends on the rate of this spinodal-like
process, which depends on the thickness of the air film. Initially,
as $H$ is increased, the air film becomes thinner, and the breakup of
the air film occurs more rapidly; thus, even though the rate of initial
drop spreading increases with $H$, the length over which the drop skates
on the air film decreases. However, as $H$ increases still further,
the thickness of the air film saturates, and hence the rate of breakup
also saturates; however, the rate of initial spreading of the drop
continues to increase with $H$. Thus, the drop always can skate over
the film of air, even as $H$ continues to increase. Interestingly, this
skating on the film of air can persist, even until $H$ increases enough
that a sheet of fluid is ejected near the expanding rim, and a splash
is produced. This suggests that dynamics of this ephemeral film of
air may be of far greater importance, and may in fact influence splashing;
however, confirmation of this speculation requires further investigation.

Acknowledgments: This work was supported by the NSF (DMR-10006546)
and the Harvard MRSEC (DMR-0820484), and the Harvard Kavli Institute
for Bio-nano-science and Technology. JMK acknowledges the support
from the NDSEG Fellowship. SMR acknowledges the support from the Yad
Hanadiv Rothschild Foundation. LM acknowledges support from the MacArthur
Foundation.

\bibliographystyle{apsrev}

\end{document}